# Water Optical Nonlinearity: Explaining Anomalously Large Electro-Optic Coefficients in Poled Silica Fibres

## JOHN CANNING

[1]*interdisciplinary Photonics Laboratories, School of Electrical & Data Engineering, University of Technology, Sydney (UTS) & School of Chemistry, The University of Sydney, Sydney NSW 2007 Australia*
*Corresponding author: john.canning@uts.edu.au*

**An explanation is offered for the large anomalous electro-optic (e.o.) effect reported by Fujiwara in 1994. It is based on the large e.o. coefficient of ordered water at an interface measured in recent years. The concept of water-based photonics is introduced, suggesting that liquid states of matter can allow ready shaping and exploitation of many processes in ways not previously considered.**

It is more than 22 years since the highest induced non-linearity within a silica ($SiO_2$) optical fibre has been reported - 5.8 pm/V when poling with 193 nm ArF exciplex emission in late 1994 [1]. UV poling had achieved an order of magnitude over anything else reported. It was groundbreaking if correct, introducing practical nonlinearity onto silicon (Si) -based platforms potentially making $LiNbO_3$ obsolete overnight. Unfortunately, at first glance the result appeared to defy the laws of physics in a material with no inversion symmetry. The failure to reproduce such results effectively consigned the work to the dustbin despite no definitive proof of any error put forward. We reconsider that work, suggesting a new photonic platform based on water.

To unravel this mystery, requires a brief review of just how symmetry in these materials can be broken. The instantaneous and parametric nonlinear optical phenomena induced within an optical fibre can, in the case where applied optical fields are not too large, be described by a Taylor series expansion of the dielectric polarization density $P(t)$ at time $t$ in terms of the applied electrical field $E(t)$:

$$P(t) = \varepsilon_0(\chi^{(1)}E(t) + \chi^{(2)}E^2(t) + \chi^{(3)}E^3(t) + \cdots) \quad (1)$$

$\chi$ represents linear and higher order nonlinear susceptibilities. The key photonic platform materials of Si, $SiO_2$ and most silicate materials are centrosymmetric (no asymmetry in the overall electric field distribution giving rise to sufficiently large polarisability that can be easily manipulated through electronic field displacement with voltage). For example, the second order term in cartesian space, $P_i(t) = 2\sum_{j,k} d_{i,j,k} E_j(t) E_k(t)$, where $d$ is the nonlinear tensor, associated with the Pockel's effect predicts a quadratic dependence on applied electric field for a traversing optical beam. For the interrogating frequency, $w$, the expression for the polarisability change with an applied field, $E_{app}$, can be reduced to [9]:

$$P(w) = \varepsilon_0(\chi^{(1)} + 2\chi^{(2)}E_{app})E(w) \quad (3)$$

From this behavior a dependence with directly applied voltage is obtained. Non-centrosymmetric materials such as $LiNbO_3$ (with $r$ coefficients between 8-30 pm/V) make use of this and the subsequent the order effects that can be generated. The material mismatch between these crystals and $SiO_2$-based photonics limits widespread deployment of photonic circuits driving attempts to create direct symmetry breaking in amorphous silicates.

There are several ways breaking symmetry in the tensor components describing polarisability. In recently reported active Si devices, apparent symmetry breaking by directly applying strain appeared to produce sizeable nonlinear coefficients [2]. The work has also been highly controversial given the strong centro-symmetric structure, the absence of any initial molecular dipole amenable to re-orientation, and the resistance to mechanical deformation of Si structure. Recently, however, such strain-induced effects have been explained as arising from the presence of SiN layers rather than the Si itself [3]. The work demonstrates the neat trick of utilizing and demonstrating optical field overlap to obtain practical device responses. These layers already have a defined local microscopic dipole that, particularly in thin film form, lends itself to deformation and alignment when significant strains are applied. The strain field has an equivalent effect to an electric field and is generally a combination of piezoelectric (negligible piezo coefficient for centro-symmetric materials) and flexo-electric contributions, the latter being of interest in understanding strain effects. The flexo-electric effect is well known in liquid crystals and more recently in solid crystals, polymers [4] and ceramics [5], including perovskites [6], potentially promising additional materials for photonics. This is because in all these materials the existing molecular dipole is already strong but randomized so there is no pre-existing macroscopic polarisation. By applying strain, randomization is removed and a macroscopic polarisability is induced giving rise, through SiN, to the observed strain gradient induced net e.o. effects in Si.

Similarly, there is a dipole arising from Si-O bonds in silica glass; however, the much more rigid constraint in tetrahedral geometry prevents similar strain-induced changes in the polarisability (in contrast to the flexibility of its longer range network). Poling, usually thermal or optical, is the method used to attempt to presumably break this tetrahedral centrosymmetry in optical fibres. It takes place in the high purity silica cladding so its rigid network geometry combined with amorphous distribution makes it difficult to break the inversion symmetry. A plausible argument is that poling has led to electro-restrictive like effects, through the attraction between the two electrodes, that have deformed the random nature of the glass through linear strains and possibly circumferential strains at the interface of the large air holes containing the attractive conductors to which large voltage is applied. Given the range of experiments looking to explore electro-restrictive effects, it is striking that this has not demonstrably generated coefficients on a scale approaching the result of Fujiwara *et al.* [1]. Electrostriction, however, did account for conventional third order nonlinearities induced by poling [7]; its contribution (on its own) may be ruled out.

A simpler approach to symmetry breaking in the material is forcing charge movement and separation through impurities or defect sites in the glass into layers orthogonal to the propagation of light. This has been the most plausible model so far utilised in the study of poled optical fibre (and other system) nonlinearities – it appears to account for the bulk of conventional poling results, potentially up to $\sim$ 1pm/V, though typically closer to 0.2 pm/V [8,9]. It was insufficient to explain the anomalously high results of 1994, which exceeded six times this value.

Charge separation under an additional field during poling requires impurities. Some of these such as $Na^+$ and $OH^-$ ions, which migrates towards the cathode and anode respectively, arise in trace amounts within the lower quality outer silica glass used in optical preform fabrication; nonetheless, despite methods of introducing more impurities the quantities and internal fields that build up appear to limit the plausibility of explaining values even approaching 1 pm/V. The use of chlorine in particular reduces $OH^-$ concentrations; $Cl^-$ itself is not typically in high concentrations. Etching experiments did confirm the presence of $OH^-$ and marked depletion layers are left behind as Na ions are visibly transported away towards the cathode [9]. It has been interesting to observe that reproducibility at 1 pm/V was not always guaranteed. A partial explanation may be the optimal process for maximising the induced field when the positive charged migrate over the fibre core [9].

So far, there has been no convincing explanation to account for the Fujiwara results. Despite irreproducibility the original authors remain adamant that the observed effects were real and that the experiments conducted with considerable care. The current author has a passing interest in this topic because he and colleague Peter Hill had converted a discarded XeCl exciplex laser into the 193 nm ArF laser for light-matter studies [10], including the Fujiwara work. There was no evidence to suggest methods were inadequate fueling the mystery around the results. Still over time, these results were dismissed whilst poling of optical fibres remains a topical theme. For example, recent results demonstrate how even a weak effect compensated by long waveguide interaction lengths can be used to generate entangled photons with entangled polarization states for quantum applications [11]. The tantalizing possibility that such demonstrations could be achieved with shorter, robust and practical lengths, even in low-loss silicate based integrated optics, arguably spurs the motivation for solving the mystery of the Fujiwara case.

Based on observations at the time the author proposes an alternative explanation, one that points to new directions and potentially new opportunities in nonlinear switching; indeed a new field. Key points any model using silica poled fibres as the benchmark has to consider include:

(1) Poling temperatures of a few hundred degrees remains way too low for any symmetry breaking in silica glasses supporting charge separation,
(2) Poling lifetimes are inconsistent with any solid phase symmetry breaking,
(3) There has been a consistent correlation with OH ions and poling magnitude reported [9], often the basis of charge separation models,
(4) The closer the core is placed to the anode channel, the higher the value of the electro-optic coefficient suggesting surface or near-surface origins,
(5) The Fujiwara results in a twin hole fibre with a separation of only $\sim$ 2 microns between channel and core, demonstrated a Bragg wavelength shift with increasing voltage applied during poling AND an increase in grating reflectivity $\Delta R \sim$ 10 dB [9].

The remoteness of the depletion and channel layer from the guiding core, $\sim$ 2 µm in the Fujiwara work [1], would necessitate large values given the low percentage overlap of the optical field in this region ($\sim$ 1 % at best); this can be minimised with appropriate fibre design and core placement but to date this has not been sufficient. On the other hand, there is one overlooked possibility related to $OH^-$ diffusion – water in the air channels supporting the electrodes. This water can be introduced both during and after the fibre is fabricated. The literature has assumed all charge separation processes are driven within the fibre cladding which appears generally true as evidenced by changes near the electrode channel.

Fig. 1. Close-up of the channel area after poling when water is present as a double layer on the surface of the glass. Hydrophilic attraction holds the layer well after bulk water affected by thermal agitation has evaporated. Figure is derived and edited from Figure 1 in [9].

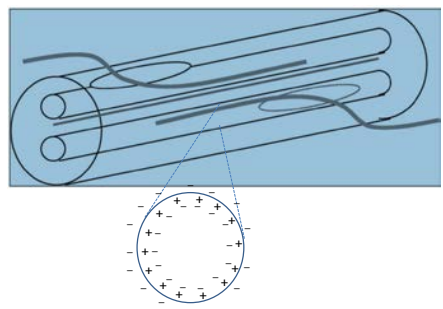

Here, the author proposes that a water layer on the channel surfaces play a key role and could help explain the anomalous result of 1994. The basis for this is that the optical fibres used by Fujiwara *et al.* were produced on a custom-made, relatively simple open-air draw tower. The presence of humidity and condensed moisture within the optical fibre channels and on the electrodes after drawing was guaranteed. Further, unlike water in bulk, poled water at interfaces *no longer has the water dipoles randomly oriented* and therefore the electro-optic coefficients of water are no longer necessarily negligible. Results reported in the literature for the electro-topic coefficients of water provide additional circumstantial evidence for this. This literature is examined to propose the case that there is instead, with poling, the formation of an electrical double layer and a possibly self-assembled, charge driven 2D water structure on the surface of the channels that is largely responsible for the symmetry breaking and charge separation in the water. This will give rise to an enormous Pockels effect *not necessarily in the silica but instead in the soft water*, potentially large enough to explain the results measured at the core of the fibre. In addition, this effect can enhance charge separation and field-build up within the fibre itself.

*The Electrical Double Layer or Equivalent*

Figure 1 shows the schematic of the fabricated device used by Fujiwara and illustrated by Fleming & An [9], copied and edited here to highlight where a water double layer would occur. An electrical double layer is a charged layer that appears at a surface when in contact with a fluid such as water. Ordinarily what triggers the formation of charges into layers at interfaces is the pre-existence or adsorption of charge on the solid interface, in this case $SiO_2$. Primarily through the dissociation of terminal silanol groups on the $SiO_2$ surface, the silica acquires negative charge. This negative charge then leads to enhanced charge separation in water into layers of alternating charges that make up the double layer. Indeed, it is well known by forcing water flow through silica channels a voltage and current can be generated, the basis of electrophoresis diagnostic methods [12]. Recent works have used this as a potential novel source of power [13,14] with sufficient voltage generated, and scalable with the number of channels [15], a method for potential energy harvesting in microfluidic or millifluidic devices. This phenomenon is present when water is present.

The double layer in the water forms and grows at a surface with an applied field - water is a dipole self-ionising to form an equilibrium with hydronium and hydroxyl ions:

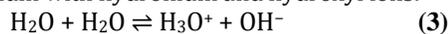
$$H_2O + H_2O \rightleftharpoons H_3O^+ + OH^- \qquad (3)$$

The local molecular structure remains enigmatic because both surface confinement and removal of charge leads to orientation and close packing of water. For condensed water in the optical fibre, pure water on silica is the primary contributor to the double layer with the proviso that this double layer potential is altered through the release of charges from the glass or electrode. A strong negative charged layer can build up further attracting impurities such as $Na^+$ as well as injecting protons into the glass. Considerable scope for optimising these effects exists.

In addition to charge separation, the density of water increases because of soft-matter dipole alignment and packing, potentially leading to molecular self-assembly expected in such confined water states. These are aligned dipoles that create charged crystal-like layers thought to possibly be similar to those reported in [16]. The interactions between them at the surface are strong enough to resist thermal motion of the water for a few layers. Regardless of the exact structural details, the net inversion symmetry of the water system is broken and all the usual rules assumed for bulk water no longer hold – in the liquid state the lack of thermal motion leads to an analogous viscous crystal-like film. Ordinarily, thermal motion prevents self-packing of the water molecules into lowest energy states and the water is random with no obvious asymmetry. But at an interface, without bonding self-packing, or self-assembly, of the molecular dipoles can occur since the interface removes the degrees of freedom for thermal motion as well as absorbing some of the local energy. The migration of proton charges away leaves a dense negative index layer. Indeed, water at an interface is consistently measured to have a higher index than that of bulk water, $n \sim (1.34\text{ -}1.39)$ [17] with scope to potentially exceed glass. This damping of thermal motion is sufficient for charge double layers to form, the thickness of which will depend on the applied fields and the quality of the surface layer. The fact that Fujiwara's result were observed with laser poling rather than thermal poling is consistent with a relatively cold, electromagnetic field process. By contrast, at the temperatures employed by thermal poling, including irradiation induced energies, this is not true for $SiO_2$. It points to new directions that can be pursued at temperatures where $SiO_2$ is similarly liquid-like.

*Gigantic Pockels Effect in Water*

Considering the layer in isolation, what is the potentially induced linear electro-optic condition in this water double layer? Nanolayers on a surface are challenging systems to measure not the least of which is that its dimensions vary depending on temperature, poling voltage and surface smoothness. The presence of such a layer with high indices may help draw an optical field if those indices can exceed silica [18]. Fortunately, the surface of an optical fibre channel will almost certainly be the smoothest possible any amorphous network can have by virtue of its production method [19]. This should maximize properties including coefficients. Despite being difficult to measure given the sort range order < 2nm, there are several reports in the literature of the induced Pockels coefficient of interfacial water on electrode surfaces being very large. On rough electrodes it has been measured as 60 up to 250 pm/V in $r_{13}$ and $r_{33}$ coefficients [17,20], exceeding $LiNBO_3$. The same group had reported earlier on transparent indium-tin oxide electrodes measurements ranging over 500 to 1700 pm/V [21] and smaller values on metallic GaN [22] consistent with oxide surfaces producing larger values – consequently, the double layer on silica would be expected to be at the higher end. Demonstrating that this is a predominantly water interface effect, the air-water interface itself has an even larger measurable Pockels effect, up to $1.54 \times 10^5$ pm/V [23]. Whilst the solid-water interfaces are attributed to both orientation of the dipoles and compression, this magnitude requires substantive orientation and structure changes at the water-air interface, consistent only with substantial movement of hydrogen away from the interface and restructuring of the water. Theoretical simulation of a 2D extended structure supports this argument [24] – one would argue that this molecular rearrangement is a primary origin of surface tension in water drops. Nevertheless, whilst the physical details remain a subject of future research, the sheer plausibility of inducing such large values using a soft material alters the potential of water and other dielectric liquids for optoelectronics and must for the basis of an important area of work.

After poling, the double layer in [17,20] was estimated to be ~ 2 nm thick with a corresponding index change > 7% ($n > 1.39$) from that of bulk water, larger than aligned water alone – this was negative for a positive bias. This magnitude is significant and both it and the thickness of the layer depends on the poling conditions applied – the likelihood that the surface roughness is larger than 2 nm in places suggests the film is less than optimal. Despite likely being an overwhelming underestimate, even with these parameters alone if the overlap integral approaches 1 %, then significant changes are possible for ~ 1000 pm/V. Both the value of the coefficient and the local refractive index could be much higher with a thicker, more tightly bound film on the smoothest of silica surfaces. Alternatively, the intense periodic 193 nm electromagnetic field used by Fujiwara can periodically pole the water – the observed increased grating strength with applied voltage supports a periodic surface variation leading to periodic variations in interfacial water properties. Add to this the one-sided nature of the grating produced by high intensity 193 nm grating writing [25] may also help to draw out the optical field towards the channel surface. However, it is also noted that an off-centre grating experiencing any mechanical effect will also produce a shift in Bragg wavelength – in the case of Fujiwara the shift to longer wavelengths would imply tensile bending. The contra-argument is that tensile bending should lead to reduced grating strength rather than increased as was observed [9].

The observation of large changes in contact angle with 193 nm irradiation of silicate glasses [26] is evidence that irradiation improves hydrophilicity or surface attraction and that this will affect interfacial properties (periodically in the Fujiwara case). In turn, enhanced dipole alignment and potentially thicker double layers accompany this.

*The Stabilty of These Changes*

What may be surprising is that the stability of layers bound to a surface can be high – suppressing thermal motion at the nanoscale charge, intermolecular attraction, which unlike chemical bonds is additive, can be strong collectively and layers of bound water can remain attached long after the bulk of water has evaporated. This has, for example, been observed for water studied on inert gold where heating is required to remove the final water from the gold surface [27]. Thus, the observed poling lifetimes exceeding 200 days or more depending on preservation and conditions beforehand, appear qualitatively consistent with a water-based material. The prospect of water as a viable new electro-optic and photonic platform material is exciting.

In conclusion, a proposed model for the origin of higher than expected electro-optic coefficients in optical fibres was presented. This model recognizes the role of interfacial water stemming both from the atmosphere during fabrication and within the glass. For conventional results, the formation of a double layer, or charged regions, in $SiO_2$ offers an alternative description of charge separation models generally, especially if the induced e.o. coefficients are sufficiently large. The ultra-smooth oxide surface may allow double layers of unmatched ordering maximising the potential Pockels effect, approaching that of air. If sufficient overlap is present then this can directly explain the results. Alternatively, the present of such fields will enhance internal fields that are also present in the fibre. Further stabilisation is possible by increasing attraction to or by the surface, perhaps utilizing laser ionization, dopants and additional charged impurities including $Na^+$ ions. Other approaches include trapping layers in nanostructure fibres to improve lifetimes to practical levels. Alternative polar liquids, including liquid crystals, may also produce large Pockels values. Properties such as response speed need further work.

Third order optical nonlinearities of water such as electric-field induced second-harmonic generation have been used previously to characterise coefficients [23,28,29] of interfacial water have not been considered. Whether or not the above model accounts for past work, the idea that water itself, particularly in ordered form, is a potentially viable and powerful nonlinear photonic medium is exciting and opens up an entirely new channel in nanophotonics and nano-electronics – the concepts apply equally to integrated photonic, electronic and microfluidic technologies including lab-in-a-fibre [30,31]. Water itself could be the optical transport medium [32] within such channels.

**Funding.** Australian Research Council (ARC) (DP140100975).